\documentclass[3p,times]{elsarticle}

\usepackage{graphicx}
\usepackage{epsfig}
\usepackage{dcolumn}
\usepackage{bbm}
\usepackage{multirow}
\usepackage[utf8]{inputenc}
\usepackage[colorlinks,linkcolor=red,anchorcolor=green,citecolor=blue,CJKbookmarks=True]{hyperref}
\usepackage[toc]{appendix}

\usepackage{mathtools}
\usepackage{color}

\usepackage{booktabs}
\usepackage[normalem]{ulem}

\begin{document}
\begin{frontmatter}
\title{$T_{cc}^{+}(3875)$ relevant $DD^*$ scattering from $N_f=2$ lattice QCD}

\author[ihep,nju]{Siyang Chen}

\author[ihep]{Chunjiang Shi\corref{cor}}
\ead{shichunjiang@ihep.ac.cn}
\author[ihep,ucas]{Ying Chen\corref{cor}}
\ead{cheny@ihep.ac.cn}

\author[ihep,ucas]{Ming Gong}

\author[ihep,ucas,chep]{Zhaofeng Liu}

\author[ihep]{Wei Sun}
\author[ihep,ucas]{Renqiang Zhang}

\cortext[cor]{Corresponding author}

\address[ihep]{Institute of High Energy Physics, Chinese Academy of Sciences, Beijing 100049, P.R. China}
\address[ucas]{School of Physics, University of Chinese Academy of Sciences, Beijing 100049, P.R. China}
\address[nju]{School of Physics, Nanjing University, Nanjing 210093, P.R. China}
\address[chep]{Center for High Energy Physics, Peking University, Beijing 100871, P. R. China}

\begin{abstract}
The $S$-wave $DD^*$ scattering in the isospin $I=0,1$ channels is studied in $N_f=2$ lattice QCD at $m_\pi\approx 350$ MeV. It is observed that the $DD^*$ interaction is repulsive in the $I=1$ channel when the $DD^*$ energy is near the $DD^*$ threshold. In contrast, the $DD^*$ interaction in the $I=0$ channel is definitely attractive 
in a wide range of the $DD^*$ energy. This is consistent with the isospin assignment $I=0$ for $T_{cc}^+(3875)$. By analyzing the components of the $DD^*$ correlation functions, it turns out that the quark diagram responsible for the different properties of $I=0,1$ $DD^*$ interactions can be understood as the charged $\rho$ meson exchange effect. This observation provides direct information on the internal dynamics of $T_{cc}^+(3875)$.
\end{abstract}

\end{frontmatter}

\section{Introduction}
Ever since the discovery of $X(3872)$ in 2003~\cite{Belle:2003nnu}, there have been quite a lot near-$D\bar{D}$ and $B\bar{B}$ threshold structures observed in experiments and are generally named $XYZ$ particles~\cite{Brambilla:2019esw}. In phenomenological studies, they are usually assigned to be conventional heavy quarkonia, $D\bar{D}$ ($B\bar{B}$) molecules, or tetraquarks. Among $XYZ$ states, $Z_c(3900)$ may be the most prominent candidate for a multiquark state since it has the minimal quark configuration $c\bar{c}u\bar{d}$ and has been observed in different experiments~\cite{BESIII:2013ris,Belle:2013yex}. Recently, LHCb reported the first doubly-charmed narrow structure $T_{cc}^+(3875)$ in the $D^0D^0\pi^+$ invariant mass spectrum, whose minimal configuration must be $cc\bar{u}\bar{d}$~\cite{LHCb:2021vvq}. The mass of $T_{cc}^+(3875)$ is measured to be below the $D^0 D^{*+}$ threshold by $-273\pm 61\pm 5^{+11}_{-14}$ keV, and its width is as small as $\Gamma=410\pm165\pm43_{-38}^{+18}$ keV (A unitarised Breit-Wigner analysis gives an even smaller width $\Gamma^U=48\pm 2_{-14}^0$ keV\cite{LHCb:2021auc}). LHCb searched other charged channels and found no evidence for the existence of a similar structure, and therefore assigned $T_{cc}^+(3875)$ to be an $I=0$ state~\cite{LHCb:2021vvq,LHCb:2021auc}. 

Prior to the observation of $T_{cc}^+(3875)$, there have been many theoretical studies on doubly-charmed tetraquarks, whose predictions of the mass and width of the ground state $J^P=1^+$ isoscalar tetraquark are consistent with those of $T_{cc}^+(3875)$\cite{Ader:1981db, Silvestre-Brac:1993zem,Semay:1994ht,Moinester:1995fk,Pepin:1996id,Gelman:2002wf,Janc:2004qn,Navarra:2007yw,Vijande:2007rf,Ebert:2007rn,Lee:2009rt,Yang:2009zzp, Carames:2011zz, Carames:2012th, Li:2012ss, Feng:2013kea,Ikeda:2013vwa,Luo:2017eub,Eichten:2017ffp,Cheung:2017tnt,Wang:2017uld,Francis:2018jyb,Maiani:2019lpu,Junnarkar:2018twb,Deng:2018kly,Liu:2019stu,Yang:2019itm,Tan:2020ldi,Lu:2020rog,Braaten:2020nwp,Cheng:2020wxa,Noh:2021lqs, Meng:2021jnw, Ling:2021bir, Albaladejo:2021vln, Du:2021zzh}. In the molecular picture, an early quark model calculation predicted the existence of a $DD^*$ bound state below the $DD^*$ threshold by $1.6\pm 1.0$ MeV~\cite{Janc:2004qn}. Recent theoretical studies find that light vector meson exchanges may induce an attractive interaction between $D$ and $D^*$~\cite{Dong:2021juy,Feijoo:2021ppq,Dong:2021bvy}. One can also refer to a recent review of the present status of theoretical studies on $T_{cc}^+(3875)$ in Ref.\cite{Chen:2022asf}. 
There are also several lattice studies performed on exotic doubly-charmed meson states. The spectra of hidden-charm and doubly charmed systems with various $J^P$ quantum numbers are explored in $N_f=2+1$ lattice QCD using meson-meson and diquark-antidiquark operators~\cite{Cheung:2017tnt}, but the results do not indicate the existence of bound states or narrow resonances, since most of lattice energy levels are close to the corresponding non-interacting meson-meson energies. Another $N_f=2+1+1$ lattice QCD study on doubly heavy tetraquarks observes the ground state of $ud\bar{c}\bar{c}$ system ~\cite{Junnarkar:2018twb}, which is below the $DD^*$ threshold by $23\pm 11$ MeV after the continuum and chiral extrapolation. However, it is not enough to claim a bound state from the lowest energy level. Apart from these studies focusing on the extraction of the finite-volume energy levels, Ref.~\cite{Padmanath:2022cvl} performs the first lattice QCD study on the pole singularity of the $DD^*$ scattering amplitude at the pion mass $m_\pi\approx 280$ MeV, and reports an $S$-wave virtual bound state pole below the $DD^*$ threshold by approximately 10 MeV, which may correspond to $T_{cc}^+(3875)$ when $m_\pi$ approaches to the physical value. 

Since the LHCb experiment observed $T_{cc}^+(3875)$ only in the $I=0$ channel, it is conceivable that the isospin-dependent interaction plays a vital role in its formation.  The existing lattice QCD studies focus on the $I=0$ channel from the point of view of tetraquark and $DD^*$ scattering, and pay little attention to the isospin-sensitive properties. Given the large negative scattering length $a=-7.16(51)$ fm of $DD^*$ scattering relevant to $T_{cc}^{+}(3875)$~\cite{LHCb:2021auc}, the characteristic size $R_a=|a|$ of $T_{cc}^+$ is too large for the present lattice QCD to investigate directly. An alternative way is to study the relevant $DD^*$ scattering in several different lattice volumes and then perform the infinite volume extrapolation to check the existence of a bound state~\cite{Beane:2003da,Beane:2008dv}. With only one lattice at hand, we cannot study the property of $T_{cc}^+$ in this way yet. We focus on the $S$-wave $DD^*$ scatterings in $I=0$ and $I=1$ channels, and explore if there are dynamical differences between them. This study may shed light on the property of the $DD^*$ interaction and provide qualitative information for future phenomenological investigations. 

This paper is organized as follows: In Section~\ref{sec:numerical} we describe the lattice setup, operator construction, and the method for studying the hadron-hadron interaction on the lattice.  The results of $DD^*$ scatterings in $I=0,1$ channels are presented in Section~\ref{sec:scattering}, and the discussions can be found in Section~\ref{sec:discussion}. Section~\ref{sec:summary} is a summary of this work.  

\section{Numerical Details}\label{sec:numerical}
\subsection{Lattice Setup}
We generate gauge configurations with $N_f=2$ degenerate $u,d$ quarks on an $L^3\times T=16^3\times 128$ anisotropic lattice. We use tadpole improved gauge action~\cite{Morningstar:1997ff,Chen:2005mg} for gluons and the tadpole improved anisotropic clover fermion action for light $u,d$ quarks~\cite{Edwards:2008ja,Sun:2017ipk}. The renormalized aspect ratio is determined to be $\xi=a_s/a_t=5.3$, and the temporal lattice spacing is set to be $a_t^{-1}=6.894(51)$ GeV \cite{Jiang:2022ffl}. Using the $a_t$ and the $\xi$, we get $a_s\approx 0.152(1)$ fm. Our bare $u,d$ quark mass parameter gives $m_\pi= 348.5(1.0)$ MeV and $m_\pi L a_s\approx 3.9$. For the valence charm quark, we adopt the clover fermion action in Ref.~\cite{CLQCD:2009nvn}, and the charm quark mass parameter is tuned to give $(m_{\eta_c}+3m_{J/\psi})/4=3069$ MeV. The distillation method~\cite{Peardon:2009gh} is used to generate the perambulators for $u,d$ quarks and the valence charm quark on our gauge ensemble. In practice, the perambulators are calculated in the Laplacian Heaviside subspace spanned by $N_\mathrm{vec}=70$ eigenvectors with the lowest eigenvalues. The parameters for the gauge ensemble are listed in Table~\ref{tab:config}.
\begin{table}[t]
	\renewcommand\arraystretch{1.5}
	\small
	\caption{Parameters of $N_f=2$ gauge ensembles with degenerate $u,d$ sea quarks.}
	\label{tab:config}
	\begin{center}
		\begin{tabular}{llllcccc}
			\toprule
			$L^3\times T$       &		$\beta$	&	$a_t^{-1}$(GeV)	&	$\xi$ 	& $N_\mathrm{cfg}$   & $m_\pi$(MeV) &	$m_{J/\psi}$(MeV) &$N_{\mathrm{vec}}$ 	\\\midrule
			$16^3 \times 128$   &   	2.0	&	$6.894(51)$	&	$ \sim 5.3$	            & $ 6950$  & $348.5(1.0)$ &	$3099(1)$	&70	  \\
			\bottomrule
		\end{tabular}
	\end{center}
\end{table}

\subsection{Operators and correlation functions}

In the lattice study of hadron-hadron scattering, one key task is to extract the lattice energy levels as precisely as possible, from which the scattering matrix elements can be parameterized with quantities reflecting the scattering properties, such as the scattering phase shift and scattering length, etc.
Concerning the properties of $T_{cc}^+(3875)$, we focus on the $DD^*$ scattering in the $J^{P}=1^+$ channel with isospin $I=0$ and $I=1$. Throughout this work, the $D(D^*)$ operators and $DD^*$ operators are built in terms of smeared quark fields. 
We use quark bilinears $O_\Gamma=\bar{q}\gamma_5 c$  for $D$ mesons and  $O_\Gamma=\bar{q}\gamma_i c$ for $D^*$ mesons (here $q$ refers to $u$ for $D^0$ and $d$ for $D^+$). 
Accordingly, the operators for $D(D^*)$ mesons moving with a spatial momentum $\vec{p}$ are obtained by the Fourier transformation $O_\Gamma(\vec{p},t)=\sum\limits_{\vec{x}}e^{-i\vec{p}\cdot\vec{x}}O_\Gamma(\vec{x},t)$.

\begin{figure}[t]
    \centering
    \includegraphics[width=0.3\linewidth]{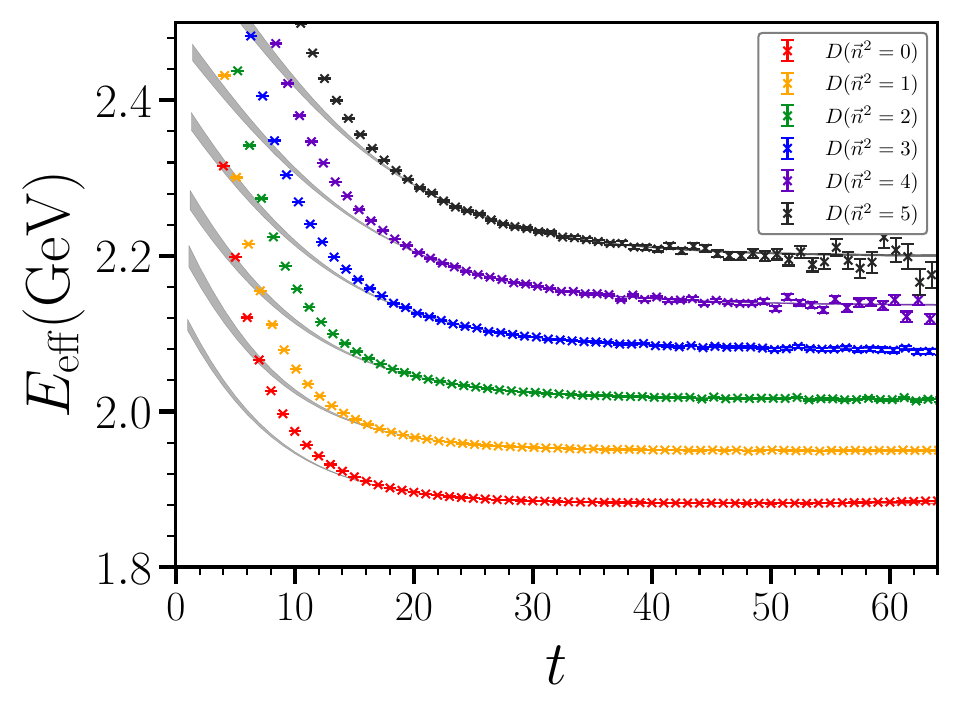}
    \includegraphics[width=0.3\linewidth]{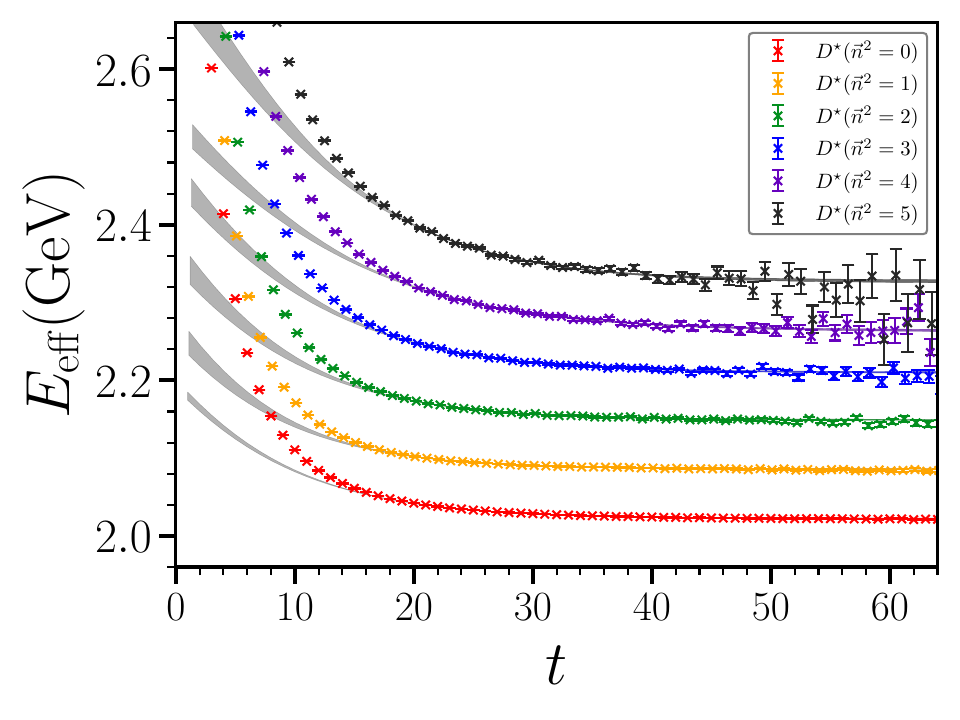}
    \includegraphics[width=0.3\linewidth]{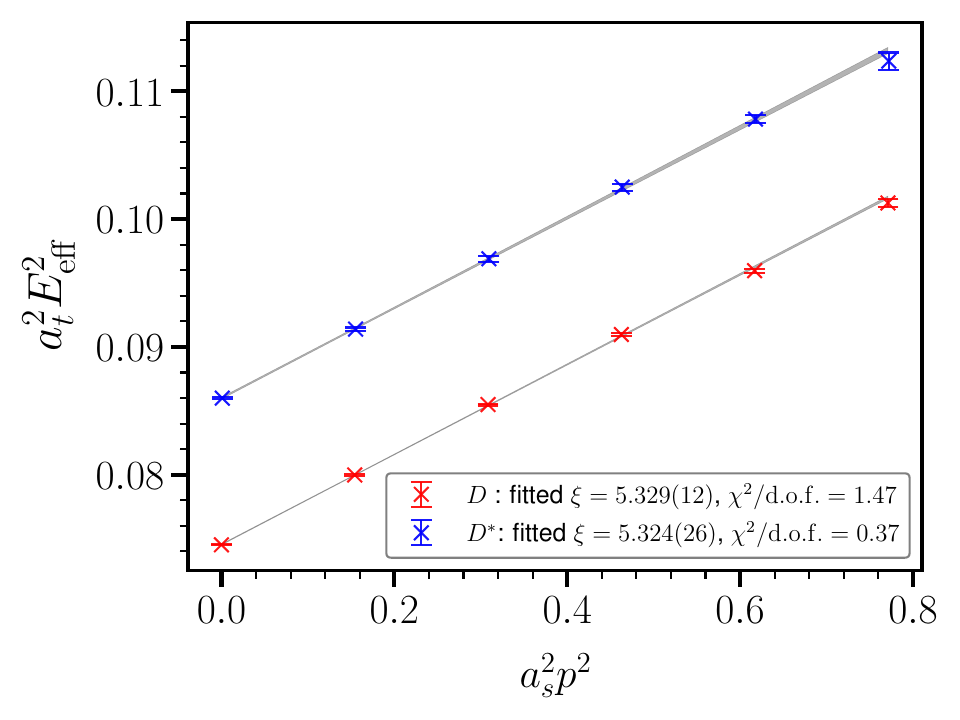}
    \caption{The effective energies and dispersion relation of $D$ and $D^*$. For the effective energies of $D$ (left panel) and $D^*$ (middle panel), the grey bands illustrate the fittings using Eq.(~\ref{eq:Dcorrelator}) in the time window $t\in[20,T-20]$. For the dispersion relations (right panel), the data points are measured energies $E_X^2(\vec{p})$ at different momenta $\vec{p}=\frac{2\pi}{a_sL} \vec{n}$ (labelled by $\vec{n}^2$) with $X$ referring to $D$ or $D^*$, and the grey bands are the fittings using Eq.~(\ref{eq:disp}).}
    \label{fig:Deff}
\end{figure}

The correlation functions of $D$ and $D^*$ at a spatial momentum $\vec{p}$ are calculated precisely using the distillation method and are parameterized as
\begin{equation}\label{eq:Dcorrelator}
    C_X(\vec{p},t)=W_1 \cosh\left[-E_X(\vec{p})(\frac{T}{2}-t)\right]+W_2\cosh\left[-E^\prime_X(\vec{p})(\frac{T}{2}-t)\right],
\end{equation}
where $X$ refers to $D$ or $D^*$ and the second term account for the higher state contamination. The modes $\vec{n}$ of the spatial momentum $\vec{p}=\frac{2\pi}{La_s}\vec{n}$ involved in this work are $\vec{n}=(0,0,0),(0,0,1),(0,1,1),(1,1,1),(0,0,2), (0,1,2)$. Fig.~\ref{fig:Deff} shows the effective energies $E_{p}^\mathrm{eff}(t)$ of $D$ (left panel) and $D^*$ (middle panel) at different momenta $\vec{p}$, which are defined by
\begin{equation}
    E_{p}^\mathrm{eff}(t) = \cosh^{-1}\frac{C_X(\vec{p},t-1)+C_X(\vec{p},t+1)}{2C_X(\vec{p},t)},
\end{equation}
and the grey bands illustrate the fit results using Eq.~(\ref{eq:Dcorrelator}) for the time interval $t\in [20,T-20]$. The results of $E_D(\vec{p})$ and $E_{D^*}(\vec{p})$ in the physical units are listed in Table~\ref{tab:Denergies} with jackknife errors. It is seen that the hyperfine splitting $\Delta m=E_{D^*}(\vec{0})-E_D(\vec{0})=139.70(57)$ MeV almost reproduces the experimental values $m_{D^{*-}}-m_{D^0}=142.0(1)$ MeV and $m_{D^{*+}}-m_{D^+}=140.6(1)$ MeV~\cite{Zyla:2020zbs}, which manifests our tuning of charm quark mass and the scale setting scheme reasonable. The momentum dependence of $E_D(\vec{p})$ and $E_{D^*}(\vec{p})$ are plotted in Fig.~\ref{fig:Deff} (right panel), where the shaded line is the fit results using the continuum dispersion relation
\begin{equation}\label{eq:disp}
    E_X^2(\vec{p})=m_X^2+\frac{1}{\xi^2} |\vec{p}|^2.
\end{equation}
The fitted $\xi$ is 5.329(12) for $D$ and 5.324(26) for $D^*$, both of which are consistent with $\xi=5.3$ in Table~\ref{tab:config}. 

\begin{table}[t]
	\renewcommand\arraystretch{1.5}
	\small
	\caption{The energies of $D$ and $D^*$ at different spatial momentum modes $\vec{n}$. The energies are converted into the values in physical units with the lattice spacing $a_t^{-1}=6.894$ GeV. }
	\label{tab:Denergies}
	\begin{center}
		\begin{tabular}{ccccccc}
			\toprule
			$\vec{n}$ modes          & $(0,0,0)$	 &	$(0,0,1)$	&	$(0,1,1)$ 	& $(1,1,1)$   & $(0,0,2)$   &	$(0,1,2)$  	\\\midrule
			$E_D(\vec{p})$(GeV)      & 1.88191(52) &1.94969(63) &2.01575(90) &2.0793(12) &2.1356(18) &2.1938(33)    \\
			$E_{D^*}(\vec{p})$(GeV)  &2.02161(91) &2.0841(15) &2.1460(25) &2.2072(30) &2.2637(33) &2.3107(72)   		  \\
			\bottomrule
		\end{tabular}
	\end{center}
\end{table}

\section{$DD^*$ scattering}\label{sec:scattering}
In this work, we only focus on the $S$-wave $DD^*$ scattering in the isospin $I=0$ and $I=1$ channels. The recent lattice study on $T_{cc}^+$ also found that the contribution of $D$-wave scattering to the $J^P=1^+$ $DD^*$ system is small enough to be neglected temporarily~\cite{Padmanath:2022cvl}.
The operators for $S$-wave $DD^*$ system with a relative $p=|\vec{p}|$ momentum can be built through 
\begin{equation}
    O_{DD^*}(p,t)=\frac{1}{N_{\vec{p}}} \sum\limits_{R\in O} O_D(R\circ\vec{p},t)O_{D^*}(-R\circ\vec{p},t),
\end{equation}
where $O_D(\vec{p},t)$ and $O_{D^*}(\vec{p},t)$ are the momentum projected single particle operators for $D$ and $D^*$, respectively, $R$ refers to the rotational operations in the lattice spatial symmetry group $O$ (the octahedral group). The operators $O_{DD^*}^{(I)}$ for a definite isospin $I$ is built according to the isospin combinations 
\begin{eqnarray}
&&I=0:~~|DD^{*}\rangle=\frac{1}{\sqrt{2}}\left(|D^0D^{*+}\rangle-|D^+ D^{*0}\rangle\right)\nonumber\\
&&I=1:~~|DD^{*}\rangle=\frac{1}{\sqrt{2}}\left(|D^0D^{*+}\rangle+|D^+ D^{*0}\rangle\right).
\end{eqnarray}

We tentatively assume the coupling between the $S$-wave the $D^*D^*$ state and $DD^*$ state is weak and do not include $D^*D^*$ operator in our calculation. 
Therefore, to  extract the energies of $DD^*$ systems, we calculate the following correlation matrix in both $I=0$ and $I=1$ channels in the framework of the distillation method, 
\begin{equation}\label{eq:DDcorr}
    C^{(I)}(p,p';t)=\frac{1}{T}\sum\limits_\tau \left\langle O_{DD^*}^{(I)}(p,t+\tau)O_{DD^*}^{(I)}(p',\tau)\right\rangle,
\end{equation}
 where we average the source time slices $\tau$ to increase the statistics. Then we solve the generalized eigenvalue problem (GEVP) 
$C^{(I)}(p,p';t) v_{p'}^{(m)}(t, t_{0})=\lambda_m(t, t_{0}) C^{(I)}(p,p';t_0) v_{p'}^{(m)}(t, t_{0})$ to get the optimized operator $O_{DD^*}^{(I)}(p_m)=v_p^{(m)}(t,t_0)O_{DD^*}^{(I)}(p)$ that couples most to the $m$-th state of $DD^*$ system with energy $E_{DD^*}^{(I)}(p_m)$. Here $p_m$ is the scattering momentum of the $m$-th state and is determined by  $E_{DD^*}^{(I)}(p_m)$ through the relation  
\begin{equation}\label{eq:scatteringmomentum}
    E_{DD^*}^{(I)}(p_m)=\sqrt{m_D^2+p_m^2}+\sqrt{m_{D^*}^2+p_m^2}.
\end{equation}
In practice, the lowest four momentum modes of $\vec{p}$ are involved in the GEVP analysis, hence the momentum modes $\vec{n}=(0,0,0)$, $(0,0,1)$, $(0,1,1)$, $(1,1,1)$ are replaced by $m=0,1,2,3$ to present the state of the $m$-th optimized operator. It is known that, under the periodic temporal boundary condition, in addition to the physical states that all the physical degrees of freedom propagate alongside in the same time direction, the so-called thermal states or wrap-around states that the $D$ and $D^*$ states propagate in opposite temporal directions~\cite{Helmes2015} also contributes to the correlation function $C^{(I)}(p,p';t)$. Therefore, the correlation function of the optimized operator $O_{DD^*}^{(I)}(p_m)$ can be parameterized as 
\begin{eqnarray}\label{eq:wrap-around}
  C^{(I)}(p_m,t) = W_1^{(I)} 
    \cosh\left(E^{(I)}_{DD^*}(p_m)(t-\frac{T}{2})\right) + W_2^{(I)}  \cosh\left(\left[E_D(p_m) - E_{D^*}(p_m)\right](t-\frac{T}{2})\right) + W^{\prime(I)} \cosh\left(E^\prime(t-\frac{T}{2})\right),
\end{eqnarray}
where the first term comes from the desired physical state, the second term accounts for the contribution of the thermal state, while the third term is introduced to account for the residual contamination from higher states. Note that $W_2^{(I)}$ is proportional to  $\exp [-(E_D(p_m)+E_{D^*}(p_m))T/2]$ which guarantees the thermal state term vanishes when $T\to \infty$. 
It turns out that this function form describes $C^{(I)}(p_m,t)$ very well in a wide time range as shown in the left panels of Fig.~\ref{fig:scattering0} and \ref{fig:scattering1}.  

 L\"{u}scher's formalism provides an approach to extracting the hadron-hadron scattering properties from the energy levels of a two-meson system in a finite box ~\cite{Luscher:1990ux,Beane:2003da}. When the energies $E_{DD^*}^{(I)}(p_m)$ is derived precisely, we can obtain the value of the scattering momentum $p_m$ using Eq.~(\ref{eq:scatteringmomentum}). Usually, one also introduces the dimensionless quantity $q=\frac{p_m L a_s}{2\pi}$ for convenience. According to L\"{u}scher's formalism, the phase shifts of $S$-wave scattering can be derived from $p$ (or $q$) by 
\begin{equation}\label{eq:phase}
    p\cot \delta_0(q^2)=\frac{2}{La_s\sqrt{\pi}}\mathcal{Z}_{00}(1,q^2)=\frac{1}{\pi L} \lim_{R\to \infty} \left[ \sum\limits_{\vec{n}\in Z_3}^{|\vec{n}|< R} \frac{1}{\vec{n}^2-q^2}-4\pi R \right].
\end{equation}
where $\mathcal{Z}_{lm}(s,q^2)$ is the L\"{u}scher zeta function \cite{Luscher:1990ux} and the second equality above is the lattice regularized version of $\mathcal{Z}_{lm}(s,q^2)$~\cite{Beane:2003da}. For the low-energy scattering, the effective range expansion (ERE) up to $\mathcal{O}(p^2)$ gives 
\begin{equation}\label{eq:ere}
    p\cot\delta_0(p)=\frac{1}{a_0}+\frac{1}{2}r_0 p^2 + \mathcal{O}(p^4)
\end{equation}
where $a_0$ and $r_0$ are the $S$-wave($l=0$) scattering length and effective range respectively. 
In the following, we discuss the $DD^*$ scatterings in the $I=0$ and $I=1$ channels in detail.

\subsection{The $I=0$ and $J^P=1^+$ $DD^*$ scattering}
\begin{figure}[t]
    \centering
    \includegraphics[width=0.3\linewidth]{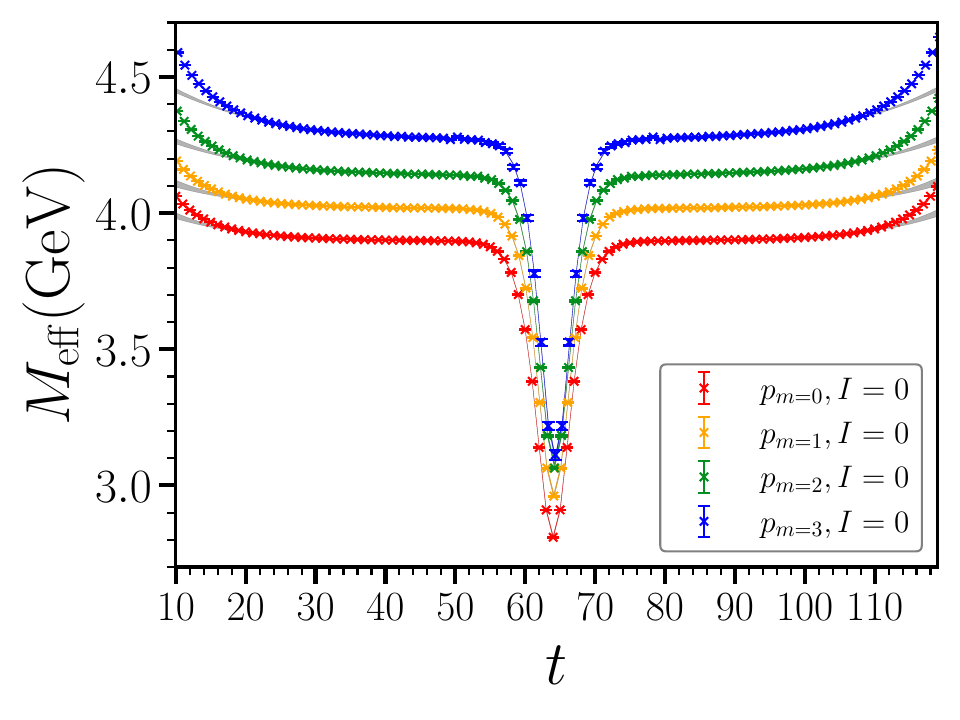}
    \includegraphics[width=0.3\linewidth]{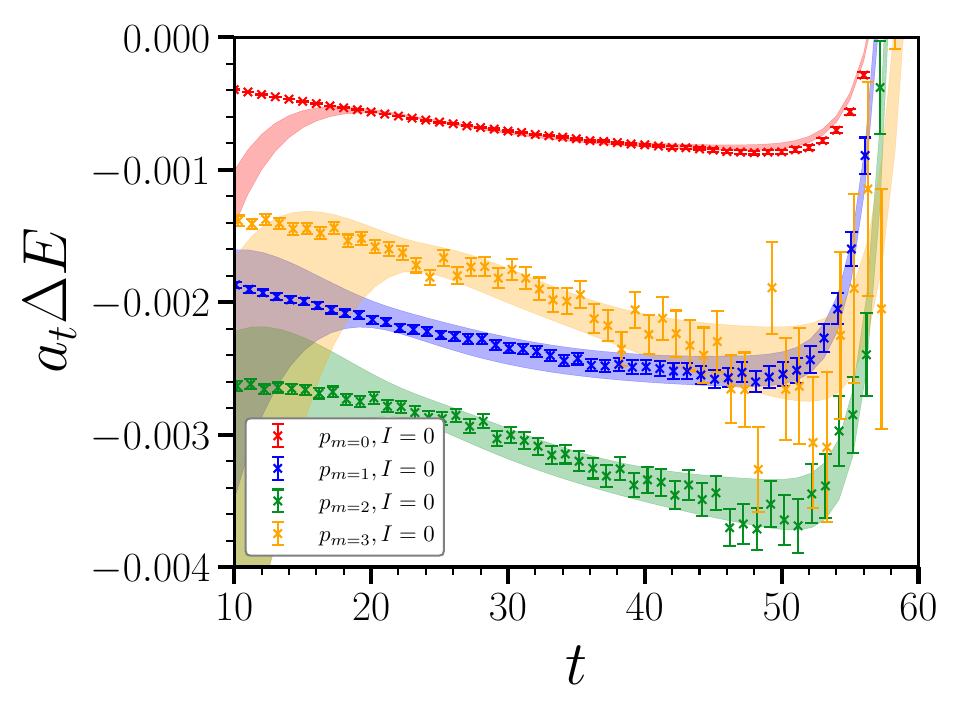}
    \includegraphics[width=0.3\linewidth]{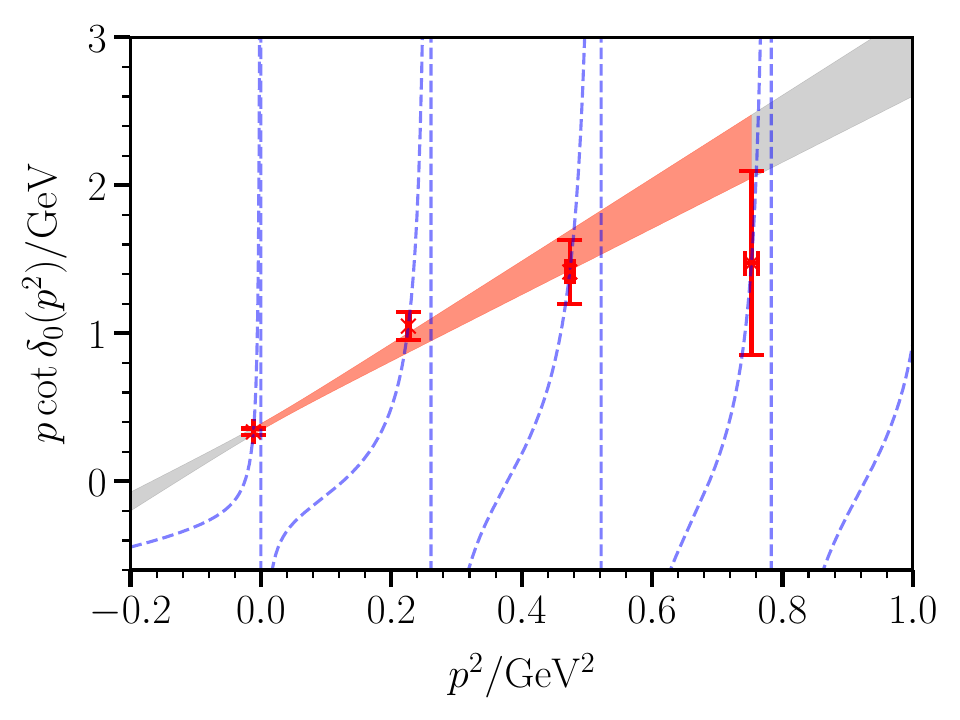}
    \caption{The results of the $DD^*(I=0)$ scattering. Left panel: Data points are the effective energies of $DD^*(I=0)$ system and the grey bands are the fits by Eq.~(\ref{eq:wrap-around}) in the time window $t\in [20,T-20]$.  Middle panel: Effective energy shifts $\Delta E(p_m,t)$ defined through the ratio function $R(p_m,t)$, where the colored bands are from the function forms of $R(p_m,t)$ defined through Eq.~(\ref{eq:Dcorrelator}) and Eq.~(\ref{eq:DDcorr}). Right panel: The phase shifts of $S$-wave $DD^*(I=0)$ scattering, where the grey band shows the result of Eq.~(\ref{eq:ere}) with best-fit parameters in Eq.~(\ref{eq:I0-best})
    and the red band illustrates the fitting range.} 
    \label{fig:scattering0}
\end{figure}
We carry out the jackknife analysis to the correlation functions $C_X(\vec{p}_m,t)$ ($X$ refers to $D$ and $D^*$) and $C^{(I)}(p_m,t)$ for all the momentum modes using equations Eq.~(\ref{eq:Dcorrelator}) and Eq.~(\ref{eq:wrap-around}), respectively (see details in Appendix~\ref{appendix:fit-details}). 
In this procedure, the energies $E_D(\vec{p}_m)$, $E_{D^*}(\vec{p}_m)$, $E_{DD^*}^{(I)}(p_m)$ for $m=0,1,2,3$ are obtained simultaneously along with the energy shifts $\Delta E^{(I)}(p_m)=E_{DD^*}^{(I)}(p_m)-E_D(\vec{p}_m)-E_{D^*}(\vec{p}_m)$ and the squared scattering momenta $p_m^2$. As shown in the left panel of Fig.~\ref{fig:scattering0} as colored bands, the function form Eq.~(\ref{eq:wrap-around}) describes $C^{(I)}(p_m,t)$ very well in the time range $t\in[20,T-20]$. The dip around $t=T/2$ also manifests the existence of the thermal states. The final results in the $I=0$ channel are listed in Table~\ref{tab:Denergies0}, where the energies with jackknife errors are converted into physical units. 
\begin{table}[t]
	\small
	\caption{The lattice results of the $S$-wave $DD^*$ scattering in $I=0$ channel. Four lowest energy levels $E_{DD^*}^{(I)}(p_m)$ corresponding to the four momentum modes are obtained. The energy shifts $\Delta E$ and the scattering momenta $p_m$ are determined accordingly. The values are in physical units converted from $a_t^{-1}=6.894$ GeV. The measured aspect 
	ratio $\xi=5.33(3)$ from the dispersion relation is used to derive the dimensionless $q^2$. All the errors here are jackknife ones. }
	\label{tab:Denergies0}
	\begin{center}
		\begin{tabular}{ccccc}
			\toprule
			$\vec{p}_m$ modes          & $m=0$	 &	$m=1$	&	$m=2$ 	& $m=3$    	\\\midrule
			$E_{D}(\vec{p}_m) + E_{D^*}(\vec{p}_m)$~(GeV)   &3.9035(14) &4.0338(19) &4.1617(29) &4.2864(36)      \\
		    $E_{DD^*}^{(I=0)}(p_m)$~(GeV)  &3.8977(14) &4.0166(15) &4.1369(18) &4.2682(28)  \\
		    \hline
			$\Delta E$~(GeV)          &-0.00582(22) &-0.0172(12) &-0.0248(23) &-0.0183(32)  \\
			$p_m^2$~(GeV$^2$)         &-0.01134(43) &0.22362(92) &0.4686(20) &0.7442(49)         \\
			$q^2=(p_m L a_s/2\pi)^2$    &-0.0440(17) &0.867(10) &1.816(22) &2.884(38) \\
			\bottomrule
		\end{tabular}
	\end{center}
\end{table}

It is seen that the energy shifts $\Delta E^{(I=0)}=E_{DD^*}^{(I=0)}(p)-E_D(p)-E_{D^*}(p)$ are uniformly negative for all the four momentum modes. This indicates the interaction between $D$ and $D^*$ in the $I=0$ channel is attractive. The energy shifts $\Delta E(p)$ are also checked through the ratio function
\begin{equation}\label{eq:ratio}
    R(p_m,t)\equiv\frac{C_{DD^*}^{(I=0)}(p_m,t)}{ C_D(\vec{p}_m,t)C_{D^*}(\vec{p}_m,t) } \sim e^{-\Delta E(p_m)t}~~(t\gg 1).
\end{equation}
This ratio function is used sometimes to estimate $\Delta E(p_m)$ from the plateau of $\Delta E(p_m,t) \equiv \ln \frac{R(p_m,t)}{R(p_m,t+1)}$. The middle panel of Fig.~\ref{fig:scattering0} shows $\Delta E(p_m,t)$ for momentum modes $m=0,1,2,3$ in $I=0$ channel, where the data points are the values from the measured correlation functions involved in Eq.~(\ref{eq:ratio}), and the colored bands illustrate the results through the function forms in Eq.~(\ref{eq:Dcorrelator}) and (\ref{eq:wrap-around}) with their fitting parameters. Obviously, $\Delta E(p_m,t)$ does not show a plateau at all, but can be well described by the function mentioned above. the slant behaviour of $\Delta E(p_m,t)$ in the intermediate time region is caused by the third term in Eq.~(\ref{eq:wrap-around}) (the excited state term), while its steep behaviour near $T/2$ is the effect of the second term (the thermal state term). This manifests that the energy shifts $\Delta E(p_m)$ listed in Table~\ref{tab:Denergies0} are derived correctly. Note that the terms for excited states in Eq.~(\ref{eq:Dcorrelator}) and (\ref{eq:DDcorr}) are necessary to describe the data. 

The scattering phase shifts $p\cot \delta_0(q^2)$ are obtained by using Eq.~(\ref{eq:phase}) at each $q^2$ and is plotted as data points in the right panel of Fig.~\ref{fig:scattering0}, where dashed lines illustrate the function form in Eq.~(\ref{eq:phase}). The fit to the four data points of lower $q^2$ using Eq.~(\ref{eq:ere}) gives   
 \begin{equation}\label{eq:I0-best}
    a_0^{(I=0)}=0.538(33)~\mathrm{fm},~~~r_0^{(I=0)}=0.99(11)~\mathrm{fm}.
\end{equation}
Our results are in line with $a_0\sim 1$ fm and $r_0\sim 1.0$ fm determined in Ref~\cite{Padmanath:2022cvl} at a lighter pion mass $m_\pi=280$ MeV. Both results indicate the attractive interaction of $DD^*$ in the $I=0$ channel. Since we have only one lattice volume, we cannot make a proper discussion on the existence of a bound state yet. 
 
\begin{table}[t]
	\small
	\caption{The lattice results of the $S$-wave $DD^*$ scattering in $I=1$ channel (similar to Table~\ref{tab:Denergies0}). }
	\label{tab:Denergies1}
	\begin{center}
		\begin{tabular}{lcccc}
			\toprule
			$\vec{p}_m$ modes          & $m=0$	 &	$m=1$	&	$m=2$ 	& $m=3$    	\\\midrule
			$E_{D}(\vec{p}_m) + E_{D^*}(\vec{p}_m)$(GeV) &3.9035(14) &4.0338(19) &4.1617(29) &4.2864(36) \\
		    $E_{DD^*}^{(I=1)}(p_m)$(GeV)   &3.9120(13) &4.0405(14) &4.1628(16) &4.2836(22)    \\\hline
			$\Delta E^{(I=1)}$(GeV)          &0.00851(23) &0.0067(12) &0.0011(23) &-0.0028(33)  \\
			$p_m$(GeV)                 &0.1289(17) &0.52131(73) &0.7226(11) &0.8815(18)    \\
			$q^2=(p_m L a_s/2\pi)^2$        &0.0644(19) &1.053(12) &2.024(24) &3.012(36) \\
			\bottomrule
		\end{tabular}
	\end{center}
\end{table}

\subsection{The $I=1$ and $J^P=1^+$ $DD^*$ scattering}

\begin{figure}[t]
    \centering
    \includegraphics[width=0.3\linewidth]{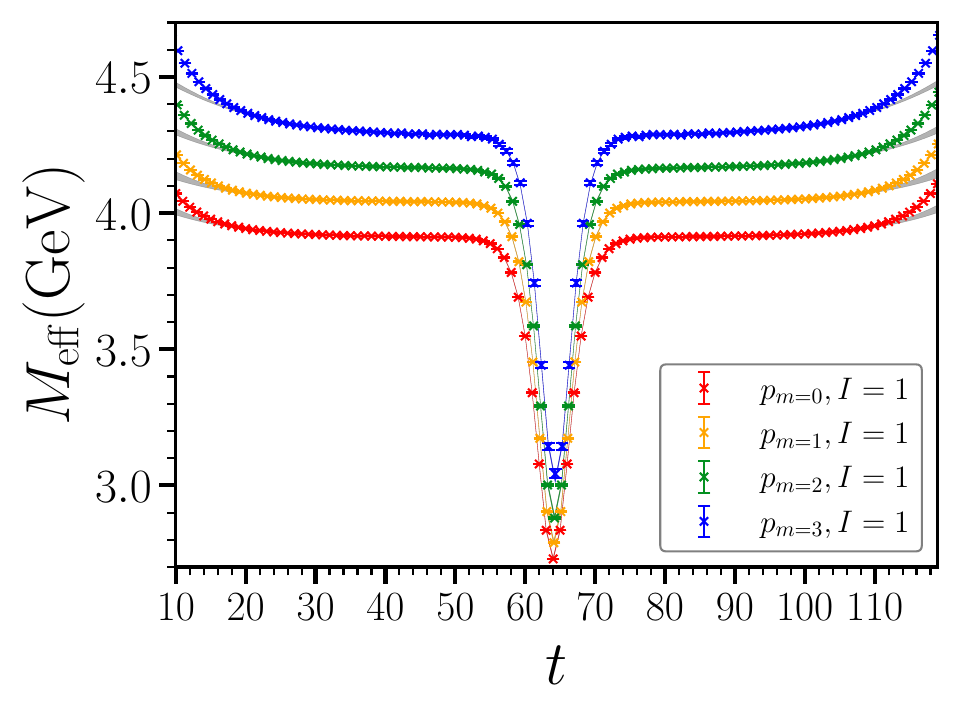}
    \includegraphics[width=0.3\linewidth]{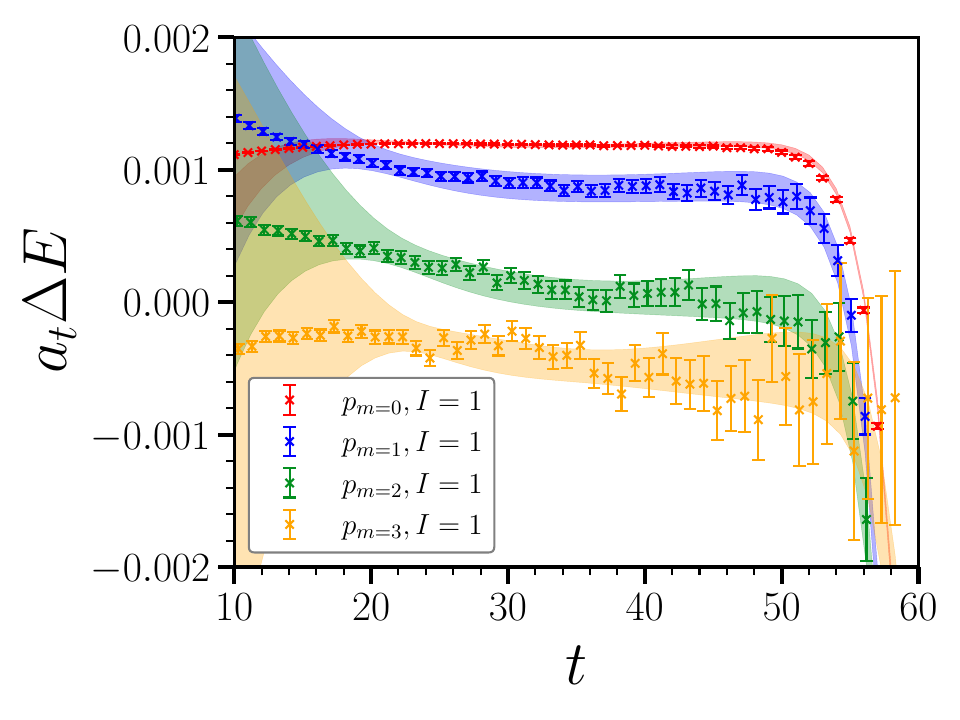}
    \includegraphics[width=0.3\linewidth]{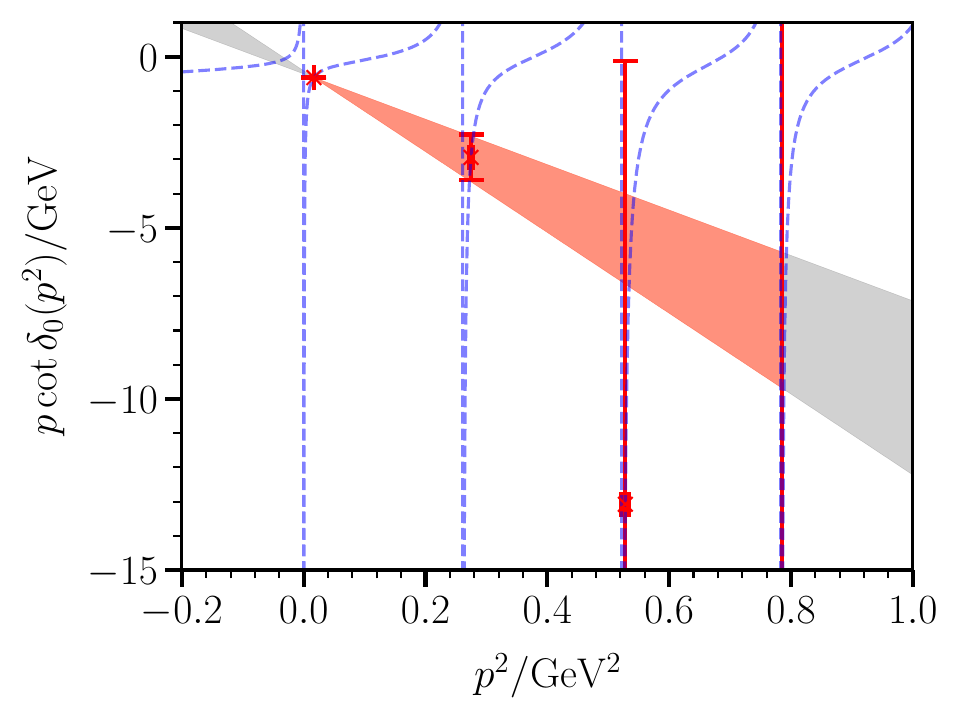}
    \caption{The results of the $DD^*(I=I)$ scattering. The three panels are similar to those of Fig.~\ref{fig:scattering0}. }
    \label{fig:scattering1}
\end{figure} 

The data analysis of the $I=1$ $DD^*$ scattering takes the same procedure as the one for the $I=0$ channel. The results of $E_{DD^*}^{(I=1)}(p_m)$ are listed in Table~\ref{tab:Denergies1} along with the 
values of corresponding energy shifts $\Delta E^{(I=1)}$, the scattering momentum $p_m$ etc.. The major results of $I=1$ $DD^*$ scattering are illustrated in Fig.\ref{fig:scattering1} similar to Fig.~\ref{fig:scattering0} for the $I=0$ case: The left panel shows the the effective energies of $C^{(I=1)}(p,t)$ and the related fits using Eq.~(\ref{eq:wrap-around}). The middle panel shows the verification of the energy shifts $\Delta E^{(I=1)}$ for different momentum $\vec{p}_m$. The right panel is for the $S$-wave phase shifts of the $DD^*(I=1)$ scattering, which is obtained from the scattering momentum $p_m$. It is seen that $E_{DD^*}^{(I=1)}(p)$ is higher than $E_{D}(\vec{p})+E_{D^*}(\vec{p})$ when it is not far from the $DD^*$ threshold (the lowest two energy levels of $E_{DD^*}^{(I=1)}(p)$). This reflects a repulsive interaction for the low-energy $D$ and $D^*$ scattering in the $I=1$ channel. When the scattering momentum $p$ is larger, the energy shifts get smaller and finally become consistent with zero within the errors. This is in striking contrast to the case of $I=0$ where the energy shifts are uniformly negative in a large range of the scattering momentum. Accordingly, the corresponding $q^2$ for the two higher energies are consistent with integers, such that when the phase shifts are determined through Eq.~(\ref{eq:phase}), their errors blow up, as shown in the right panel of Fig.~\ref{fig:scattering1}. The fit to these phase shifts using Eq.~(\ref{eq:ere}) gives the scattering length and the effective range 
as  
\begin{equation}
    a_0^{(I=1)}=-0.433(43)~\mathrm{fm},~~~r_0^{(I=1)}=-3.6(1.0)~\mathrm{fm}.
\end{equation}

\section{Discussion}\label{sec:discussion}
\begin{figure}[t]
    \centering
    \includegraphics[width=0.45\linewidth]{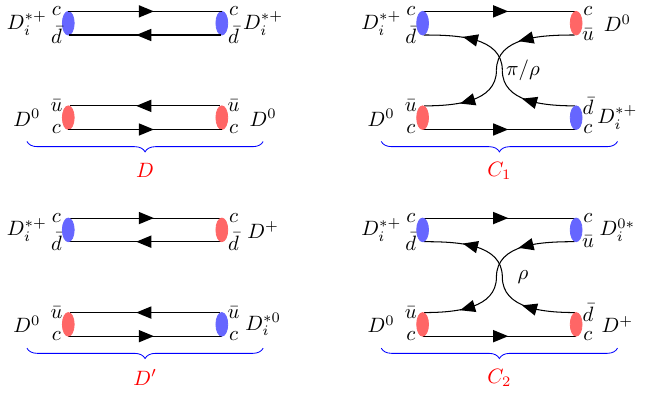}
    \includegraphics[width=0.35\linewidth]{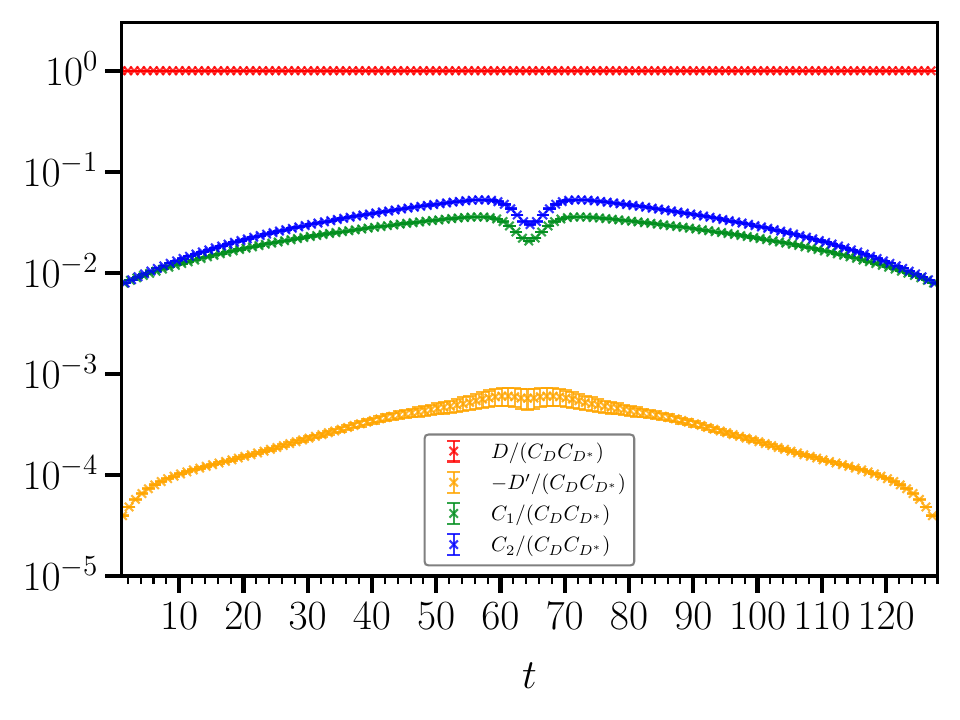}
    \caption{The components of the correlation function $C^{(I)}(p,t)$. Left panel: The schematic quark diagrams of the four terms $D$, $C_1(\pi/\rho)$, $D'$ and $C_2(\rho)$ that contribute to $C^{(I)}(p,t)$. Right panel: The relative magnitudes of the four terms for the case of $\vec{p}=0$, which are scaled by the $C_D(\vec{p}=0,t)C_{D^*}(\vec{p}=0,t)$.}
    \label{fig:contraction}
\end{figure}

In the previous section, we present the numerical results of the $S$-wave $DD^*$ scattering in the $I=0$ and $I=1$ channels. The major observation is that $DD^*$ interaction is 
attractive for $I=0$ in a wide momentum range and repulsive for $I=1$ when the energy of $DD^*$ is near the $DD^*$ mass threshold. It is conceptually in agreement with the observation of LHCb~\cite{LHCb:2021vvq} that the $T_{cc}^+$ state is found only in the  $D^0 D^{*+}$ system.

To understand the isospin-dependent interaction of $DD^*$, let us take a closer look at the quark diagrams (after the Wick contraction) which contribute to the correlation functions $C^{(I)}(p,t)$. There are four distinct terms whose schematic quark diagrams are shown in the left part of Fig.~\ref{fig:contraction}:  The diagram on the upper left side is named $D$ (direct) term which comes from the direct contractions between $O_D (O_{D^*})$ in the sink and source operators. The diagram on the upper right side is called the $C_1(\pi/\rho)$ (crossing) term which involves either the $u,d$ quark exchange effects (as illustrated in the figure) or charm quark exchange (if flipping upside down the positions of $D^0$ and $D_j^{*+}$ on the right-hand side). In the lower-left diagram, $D'$ is the direct contraction between $D$ and $D^*$. The lower right diagram $C_2(\rho)$ also illustrates a $u,d$ quark exchange one. As such $C^{(I)}(p,t)$ can be abbreviated as 
\begin{equation}
    C^{(I)}(p,t) = D - C_1(\pi/\rho) + (-)^{I+1} \left( D' - C_2(\rho) \right),
\end{equation}
where the minus signs of $C$ terms come from the single quark loops after Wick contraction. 

 The contributions of these terms to $C^{(I)}(p,t)$ at $\vec{p}=0$ are checked to have the hierarchy $D \gg C_2( \rho) \gtrsim C_1(\pi/\rho)\gg D'$ with each level being smaller by roughly two orders of magnitude, as shown in the right panel of Fig.~\ref{fig:contraction}, where the magnitudes of $D$, $C_1(\pi/\rho)$, $C_2(\rho)$ and $D'$ at $\vec{p}=0$ are scaled by the product of single meson correlation functions $C_D(\vec{p}=0,t)$ and $C_{D^*}(\vec{p}=0,t)$ (abbreviated by $C_DC_{D^*}$). The contribution of $D'$ term is quite small and negligible in the following discussion. The $C_1(\pi/\rho)$ term contributes equally to $C^{(I=0)}(p,t)$ and $C^{(I=1)}(p,t)$, while the contributions of $C_2(\rho)$ have opposite signs for $I=0,1$ and are necessarily responsible for the energy difference of $E_{DD^*}^{(I=0,1)}(p)$. As shown in the right panel of Fig.~\ref{fig:contraction}, in the intermediate time range, $C_1(\pi/\rho)/(C_DC_{D^*})$ and $C_2(\rho)/(C_DC_{D^*})$ show approximately linear behaviors in the logarithmic scale with positive slopes, while $D/(C_DC_{D^*})$ is almost a flat line throughout the time range. Since $C_DC_{D^*}$ behaves as $We^{-(m_D+m_{D^*})t}$ in the intermediate time range, the $D$ term must have a similar time dependence, namely, $A_0 e^{-E_0 t}$ with $E_0\approx m_D+m_{D^*}$. Accordingly, the time dependence of $C_1(\pi/\rho)$ and $C_2(\rho)$ is also approximately exponential, and can be expressed qualitatively as $A_0\epsilon_i e^{-E_i t}$,where $i=1,2$ refer to $C_1(\pi/\rho)$ and $C_2(\rho)$, respectively, and $\epsilon_i\sim \mathcal{O}(10^{-2})\ll 1$ is indicated by the figure. On the other hand, the positive slopes of $C_1(\pi/\rho)/(C_DC_{D^*})$ and $C_2(\rho)/(C_DC_{D^*})$ imply that $C_1(\pi/\rho)$ and $C_2(\rho)$ damp in time more slowly than $D$ does, such that one has $E_0-E_i=\delta E_i>0$. Thus one has the approximation (see Appendix~\ref{appendix:compositness}) for the energy of the $DD^*$ system 
 \begin{equation}\label{eq:EIdiff}
    E_{DD^*}^{(I)}\approx \ln \frac{C^{(I)}(p,t)}{C^{(I)}(p,t+1)}\approx E_0 + \epsilon_1 \delta E_1 e^{\delta E_1 t}+ (-)^{I+1} \epsilon_2 \delta E_2 e^{\delta E_2 t}
\end{equation}
in the time range $t\in [20,50]$ where $\delta E_i t\ll 1$. The second term on the right-hand side of Eq.~(\ref{eq:EIdiff}) comes from the $C_1(\pi/\rho)$ contribution and is positive for both $I=0,1$ channels. This means the $C_1(\pi/\rho)$ term reflects a repulsive interaction. In contrast, the third term, which is contributed from $C_2(\rho)$, is positive for $I=1$ and negative for $I=0$, and thereby manifests a repulsive interaction for $I=1$ and an attractive interaction for $I=0$. On the other hand, as shown in the right panel of Fig.~\ref{fig:contraction}, the curve for $C_2(\rho)$ is uniformly higher than that for $C_1(\pi/\rho)$ and thereby implies $\epsilon_2\gtrsim\epsilon_1$. In the meantime, the larger slope of $C_2(\rho)$ indicates $\delta E_2>\delta E_1$, such that one has $\epsilon_1 \delta E_1 e^{\delta E_1 t} < \epsilon_2 \delta E_2 e^{\delta E_2 t}$. In other words, the combined effects of the $C_1(\pi/\rho)$ and $C_2(\rho)$ contribution result in negative energy shifts from the non-interacting $DD^*$ energy $E_D(p)+E_{D^*}(p)$, which reflects the totally attractive interaction between $D$ and $D^*$ in the $S$-wave $I=0$ channel. One can see Appendix~\ref{appendix:compositness} for itemized information.

On the hadron level, the four terms depicted in Fig.~\ref{fig:contraction} can be interpreted as follows:
\begin{itemize}
    \item $D$ term: It involves two separately closed quark diagrams, each of which is the propagator of $D(D^*)$ meson. After the gauge averaging, the two parts can have an interaction mediated by at least two gluons that are necessarily in a color singlet. Intuitively, quarks frequently exchange gluons among themselves during their propagation. The ``motion" status of light quarks can be changed more easily by absorbing or emitting a (not hard) gluon, such that their trajectories in the spacetime are zigzag and may develop meson-exchange interactions, such as exchanges of $\sigma$, $\omega$, etc., on the hadron level. Either gluon exchanges on the quark level or meson exchanges on the hadron level, the resultant effects are very tiny since the contribution of this term is very close in magnitude (after the subtraction of the contribution from the wrap-around states) to the product of the correlation functions of single $D$ and $D^*$ mesons. 

    \item $D'$ term: This also involves two closed quark diagrams, however, each one connects two different mesons $D$ and $D^*$. This diagram contributes to $C_{DD^*}(p,t)$ only when color singlet gluon exchanges (at least two gluons also) take place between the two parts after the gauge average. On the hadron level, the interaction can be mediated by $\eta, \omega$, etc. However, empirically in our study, it is found these effects are very weak, and the contribution from the $D'$ term is negligible in comparison with the other terms.   
   \item $C_1(\pi/\rho)$ term: As shown in the right upper part of Fig.~\ref{fig:contraction}, there are explicit $u,d$ quark exchanges between $D$ and $D^*$ during their temporal propagation. This exchange effect can be viewed as that of the charged meson ($\pi^\pm$,$\rho^\pm$, etc.) on the hadron level. If we flip the positions of $D^0$ and $D^{*+}$ on the right-hand side, the figure implies a $c\bar{c}$ exchange process, and accordingly charmonium $V_c$ ($J/\psi$, $\psi'$, etc.) exchange process on the hadron level. Since $C_1(\pi/\rho)$ contributes equally to $C_{DD^*}^{(I=0)}(p,t)$ and $C_{DD^*}^{(I=1)}(p,t)$, according to our discussion above, these intermediate meson exchanges on the hadron level result in a repulsive interaction to the $DD^*$ system. Note that vector meson exchange models~\cite{Feijoo:2021ppq,Dong:2021bvy} also obtain a repulsive interaction for the $J/\psi$ exchange.  
    \item $C_2(\rho)$: This term also comes from the $u,d$ quark exchanges. On the hadron level, since the $\mathcal{P}$-parity conservation prohibits the $DD\pi$ interaction, the effect of light quark exchange can be reflected mainly by the charged $\rho$ exchange, which provides an attractive interaction for the $S$-wave $I=0$ $DD^*$ system and a repulsive interaction for the $S$-wave $I=1$ $DD^*$ system. Furthermore, the observation $E_{DD^*}^{(I=0)}(p)<E_D(p)+E_{D^*}(p)$ indicates that this attractive $\rho$-exchange effect overcomes the repulsive interaction reflected by $C_1(\pi/\rho)$ term and results in a total attraction interaction. This result is in qualitative agreement with those in Refs.~\cite{Dong:2021juy,Feijoo:2021ppq,Dong:2021bvy}.
\end{itemize}

\section{Summary}\label{sec:summary}
The $S$-wave $DD^*$ scattering are investigated from $N_f=2$ lattice QCD calculations on a lattice with $m_\pi\approx 350$ MeV and $m_\pi L a_s\approx 3.9$. Benefited from the large statistics, several lowest energy levels of the $DD^*$s of isospin $I=0$ and $I=1$ are determined precisely through the distillation method and by solving the relevant generalized eigenvalue problems. In the $I=1$ case, the $DD^*$ energy $E_{DD^*}^{(I=1)}(p)$ is higher than the corresponding non-interacting $DD^*$ energy $E_D(\vec{p})+E_{D^*}(\vec{p})$ threshold, and manifests a repulsive interaction between $D$ and $D^*$. But when the scattering momentum $p$ rises large, the difference of  $E_{DD^*}^{(I=1)}(p)$ and $E_D(\vec{p})+E_{D^*}(\vec{p})$ becomes smaller and even indiscernible. In the $I=0$ case, the $DD^*$ energy $E_{DD^*}^{(I=0)}(p)$ is uniformly lower than $E_D(\vec{p})+E_{D^*}(\vec{p})$ when $p$ goes up to around $800$ MeV, and reflects definitely an attractive interaction between $D$ and $D^*$ in the $I=0$ state. It is consistent with the experimental assignment $I=0$ for $T_{cc}^+(3875)$ given a $DD^*$ bound state. Based on these energy levels, the $S$-wave phase shifts of $DD^*$ scattering in $I=0,1$ channels are derived using L\"{u}scher's finite volume formalism.  The effective range expansions give the following  scattering lengths $a_0^{(I=0,1)}$ and the effective ranges $r_0^{(I=0,1)}$ 
\begin{eqnarray}
    a_0^{(I=0)}&=&~0.538(33)~\mathrm{fm},~~~r_0^{(I=0)}=0.99(11)~\mathrm{fm},\nonumber\\
    a_0^{(I=1)}&=&-0.433(43)~\mathrm{fm},~~~r_0^{(I=1)}=-3.6(1.0)~\mathrm{fm}.
\end{eqnarray}

To understand the isospin dependence of the $DD^*$ interaction, further analysis is performed on the components of $DD^*$ correlation functions. It is found that the difference between the $I=0$ and $I=1$ $DD^*$ correlation functions comes mainly from the $C_2(\rho)$ term  that $D$ and $D^*$ exchange $u,d$ quarks when propagating in the time direction. 
This term can be viewed as the charged vector $\rho$ meson exchange in the hadron level and contributes to the $I=0$ and $I=1$ $DD^*$ correlation functions with opposite signs. As a result, it raises the $DD^*$ energy in the $I=1$ channel, and pulls it down in the $I=0$ channel. This provides a shred of strong evidence that the $DD^*$ interaction induced by the charged $\rho$ meson exchange may play a crucial role in the formation of $T_{cc}^+(3875)$. This is in qualitative agreement with the results of recent phenomenological studies~\cite{Dong:2021juy,Feijoo:2021ppq,Dong:2021bvy}.

\vspace{0.5cm}
\section*{Acknowledgements} 
We thank Prof. Q. Zhao of IHEP for valuable discussions. This work is supported by the Strategic Priority Research Program of Chinese Academy of Sciences (No. XDB34030302), the National Key Research and Development Program of China (No. 2020YFA0406400) and the National Natural Science Foundation of China (NNSFC) under Grants No.11935017, No.11775229, No.12075253, No.12070131001 (CRC 110 by DFG and NNSFC), No.12175063. The Chroma software system~\cite{Edwards:2004sx} and QUDA library~\cite{Clark:2009wm,Babich:2011np} are acknowledged. The computations were performed on the HPC clusters at the Institute of High Energy Physics (Beijing) and China Spallation Neutron Source (Dongguan), and the CAS Sunrise-1 computing environment.

\bibliographystyle{elsarticle-num}
\bibliography{ref}

\section*{Appendix}
\begin{appendices}

\section{Additional information for the dispersion relation of $D$ and $D^*$}\label{appendix:dispersion}
  \begin{figure}[!htbp]\label{fig:phase_shifts_GeV}
        \centering
        \includegraphics[width=0.45\linewidth]{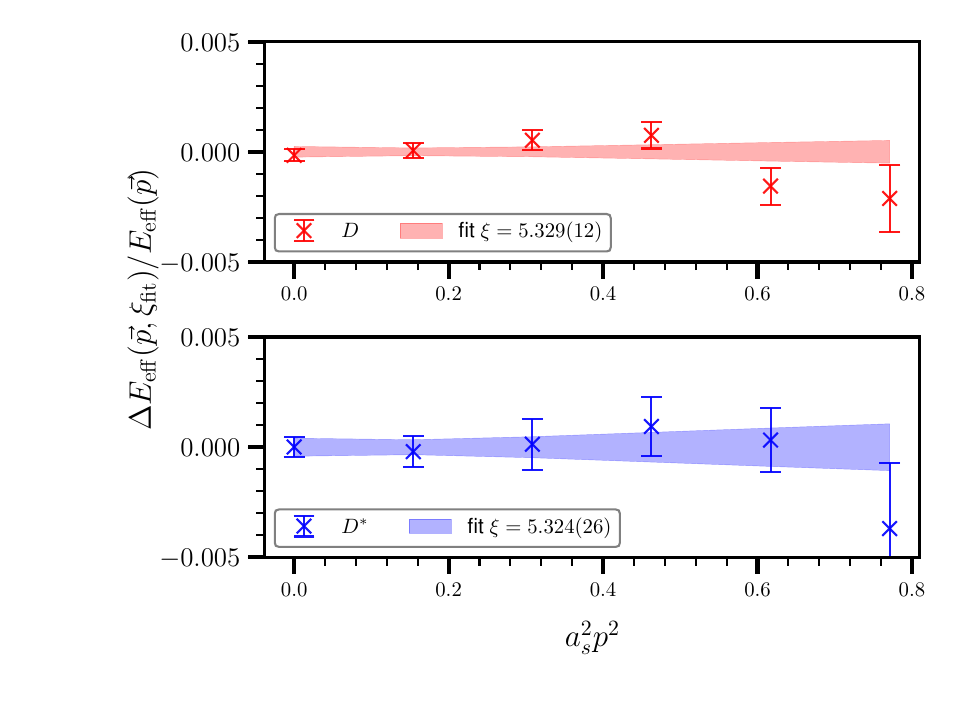}
        \caption{The relative deviation of the fit to dispersion relation. $\Delta E \equiv  E_{\mathrm{eff}}(\vec{p}) - \bar{E}_{\mathrm{cont}}(\vec{p},\xi_{\mathrm{fit}})$, the colored bands present fitting errors. }
\end{figure}

\section{Grouped jackknife analysis}\label{appendix:fit-details}
\begin{figure}[!htbp]
        \centering
        \includegraphics[width=0.45\linewidth]{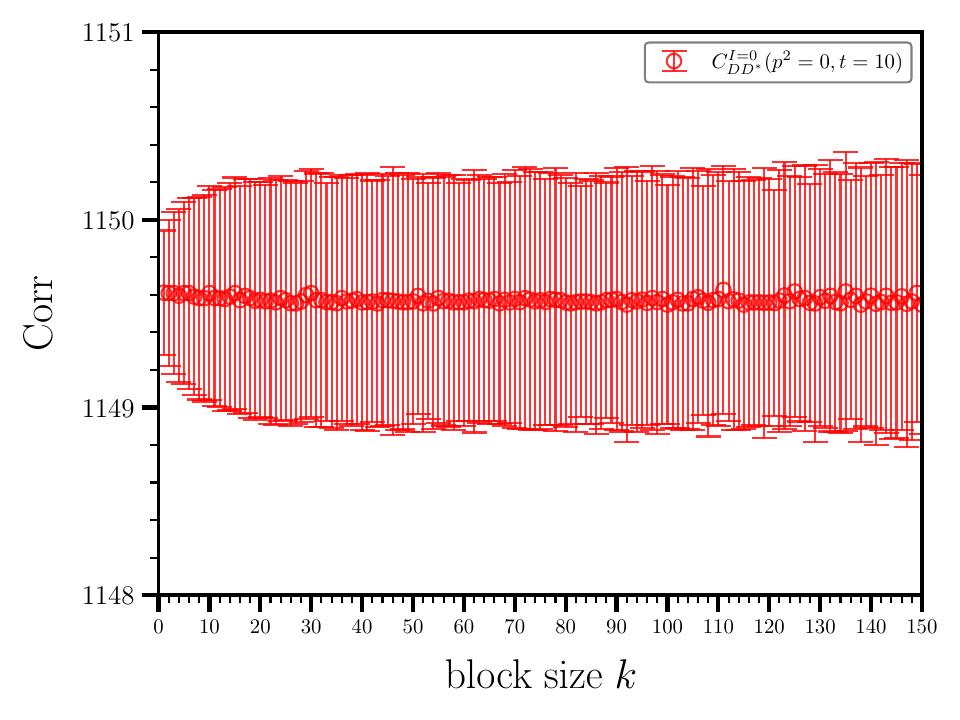}
        \includegraphics[width=0.45\linewidth]{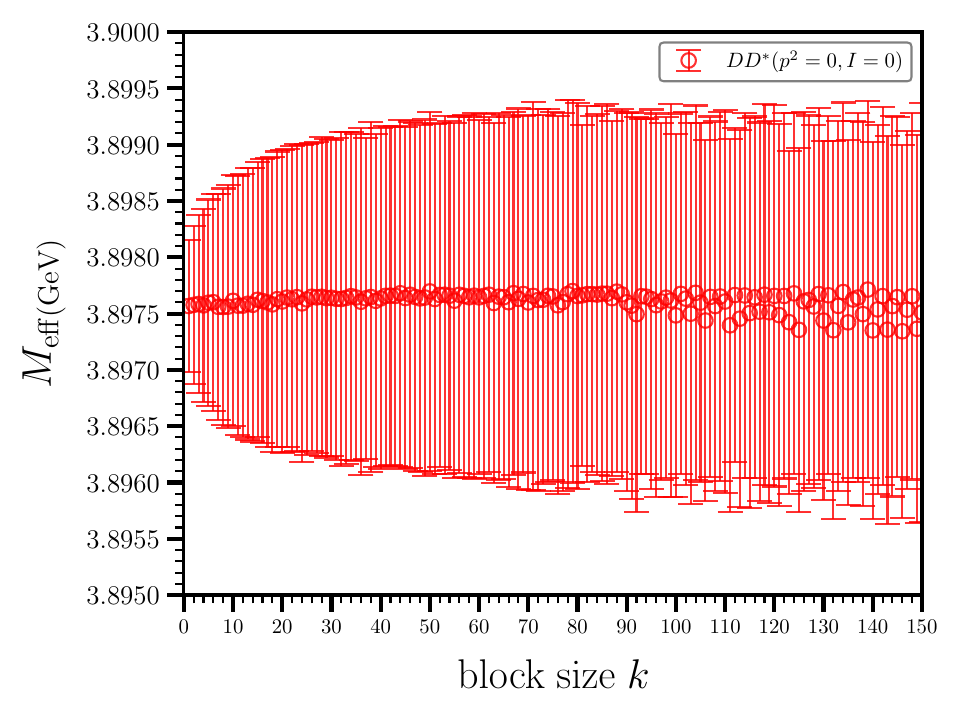}
        \caption{The statistical errors with the size $k$ in each block.}
        \label{fig:bins-error}
    \end{figure}

    Since the gauge configurations are generated through a Markov chain, they are not completely independent. Therefore, the jackknife resampling technique is utilized in our data analysis procedure. 
    We group the $N$ configurations into $n$ blocks with each block including $k=N/n$ configurations. 
    The measurements in each block are averaged as one individual measurement, then the one-eliminating jackknife analysis is performed. 
    We vary the size $k$ of the block to check the $k$-dependence of the statistical errors and find that the errors of observables increase gradually when $k$ increases and finally saturate beyond some value of $k$. Fig.~\ref{fig:bins-error} shows the $k$-dependence of the errors of the correlation function $C_{DD^*}^{(I=0)}(t)$ and the $DD^*$ energy $E_{DD^*}$. It is obvious that the errors saturate when $k\gtrsim 50$, eventually we choose block size $k=50$ for our measurements to avoid the underestimation of statistical errors. 
    
\section{Components of $DD^*$ energy $E_{DD^*}^{(I)}$}\label{appendix:compositness}
   We have the following major observation from the right panel of Fig.~\ref{fig:contraction}: 
    \begin{itemize}
        \item The $y$-axis is plotted in the logarithmic scale, and $D$ term (denoted by $D(t)$ here), $C_1(\pi/\rho)$ term ($C_1(t)$), $C_2(\rho)$ term ($C_2(t)$) and $D'$ term are scaled by $C_{D}(t)C_{D^*}(t)$;
        \item $D(t)/(C_{D}(t)C_{D^*}(t))$ is almost flat and can be described as $Ae^{-E_0 t}$ with $E_0\approx m_D+m_{D^*}$ (note that we discuss the $\vec{p}=0$ case). 
        \item The linear rising behaviors of $C_{1,2}(t)/(C_{D}(t)C_{D^*}(t))$ in the intermediate time range imply that 
        \begin{equation}
            C_i(t)\approx A\epsilon_i e^{-(E_0-\delta E_i)t}
        \end{equation}
        with $0<\epsilon_i \sim \mathcal{O}(10^{-2})\ll 1$ and $\delta E_i >0$ ($C_i(t)$ is approximately $\mathcal{O}(10^{-2})$ of magnitude smaller then $D(t)$. 
        \item $C_2(t)$ is uniformly higher than $C_1(t)$ and has a large slope. This implies $\epsilon_2>\epsilon_1$ and $\delta E_2>\Delta E_1$.
        \item $D'$ term is approximately $\mathcal{O}(10^{-4})$ of magnitude smaller than $D$ term and is ignored in the discussion. 
    \end{itemize}
    
    Based on these observations, in the intermediate time region, the energy of $DD^*$ can be estimated as  

    \begin{eqnarray}
            E_{D D^{*}}^{(I)} &\approx& \ln \frac{C^{(I)}(p, t)}{C^{(I)}(p, t+1)}= \ln \frac{D(t) - C_1(t) - (-)^{I+1} C_2(t)}{D(t+1) - C_1(t+1) - (-)^{I+1} C_2(t+1)} \nonumber\\
            &=&\ln \frac{D(t)}{D(t+1)}+\ln \frac{ 1- \epsilon_1 \exp[\delta E_1 t] - \epsilon_2 (-)^{I+1} \exp[\delta E_2 t]}{1 - \epsilon_1 \exp[\delta E_1 (t+1)] - \epsilon_2(-)^{I+1} \exp[\delta E_2 (t+1)]} \nonumber\\
            &\approx& E_0 + \epsilon_1 \exp( \delta E_1 t)(e^{\delta E_1}-1) + \epsilon_2 (-)^{I+1}\exp(\delta E_2 t) {(e^{\delta E_2}-1)} \nonumber\\
            &\approx& E_{0}+\epsilon_{1} \delta E_{1} e^{\delta E_{1} t}+(-)^{I+1} \epsilon_{2} \delta E_{2} e^{\delta E_{2} t}
    \end{eqnarray}
    where $\epsilon_i, \delta E_i t\ll 1$ is used. 
    
\end{appendices}

\end{document}